\documentstyle [12pt]{article}
\def\double{\baselineskip=24pt}

\def\arcmin{{$^{\prime}~$}}
\def\ksec{\rm {km~s$^{-1}$~}}

\def\ref{\par\noindent\hangindent 20pt}
\begin{document}

\vskip 1.5cm

\titlepage

\begin{center}
{\bf SUBCLUSTER MERGERS AND GALAXY INFALL IN A2151}
\vskip 1.2cm
Christina M. Bird \\
{\it Department of Physics and Astronomy \\
University of Kansas\\
Lawrence, KS 66045-2151} \\
Int:  tbird@kula.phsx.ukans.edu\\
\medskip
David S. Davis \\
{\it Department of Astronomy\\
University of Maryland\\
College Park, MD 20742}\\
Int:  ddavis@attilla.gsfc.nasa.gov\\
\medskip
\& \\
\medskip
Timothy C. Beers \\
{\it Department of Physics and Astronomy\\
Michigan State University\\
East Lansing, MI 48824}\\
Int:  beers@msupa.pa.msu.edu\\
\end{center}

\medskip

\begin{abstract}

We have obtained a 12.5 ksec image of the Hercules Cluster, A2151,
with the {\it ROSAT} PSPC.
Comparison of the optical and X-ray data suggest the presence of at
least three distinct subclusters in A2151.  The brightest X-ray emission
coincides with the highest-density peak in the galaxy distribution, and
is bimodal.
The northern subclump, distinct in position and velocity, has {\it no}
detectable X-ray gas.  The eastern subclump, apparent in the optical
contour map, is indistinguishable from the main clump in velocity space but
is clearly visible in the X-ray image.

X-ray spectra derived from the central peak of emission yield a best-fit
temperature of 1.6 keV.  The emission coincident with the eastern clump
of galaxies is cooler, 0.8 keV, and is outside the 90\% confidence intervals
of the central peak temperature.

We suggest that the eastern and central subclusters have recently
undergone a merger event.  The lack of X-ray emission to the north
suggests that those galaxies do not form a physically-distinct
structure (i.e. they are not located within a distinct gravitational
potential), but rather that they are falling into the cluster core
along the filament defined by the Hercules Supercluster.
\end{abstract}

\medskip

\noindent{Submitted to {\it The Astronomical Journal} 4 March 1994}

\bigskip
\noindent
{\it Subject headings:}  galaxies:  clusters:  individual:  A2151 ---
galaxies:  clustering

\vfill

\newpage

\double

\begin{center}
{\bf 1. INTRODUCTION}
\end{center}

The early assumption that most clusters of galaxies are relaxed has proven
to be incorrect.  X-ray imaging observations with the {\it Einstein}
satellite first revealed that complexity in the intracluster medium (ICM)
is frequent (Forman et al.\ 1981;  Henry et al.\ 1981), in contrast to
the smooth configurations assumed in early studies (cf.\ Kent \& Gunn 1982;
Kent \& Sargent 1983).  Since then, both
optical (Geller \& Beers 1982; Baier 1983; Beers et al.\ 1991; Bird 1993,
1994a,b) and X-ray studies (Jones \& Forman 1992; Davis \& Mushotzky 1993;
Mohr, Fabricant \& Geller 1993) have revealed that many, if not all, clusters
possess significant deviations from an equilibrium state.  Peculiar
velocities of brightest cluster galaxies, in particular the ultraluminous cD
galaxies once thought themselves to be signatures of advanced dynamical
evolution, are correlated with the presence of substructure (Bird 1994b).
The properties of radio galaxies are dependent on their local physical
environments, not their global cluster properties (Burns et al.\ 1994).
Observations of clusters in all wavebands reveal that dynamically-relaxed,
spherical systems seem to be the exception rather than the norm.

In many massive clusters of galaxies, the morphologies revealed by either
galaxies or gas are similar.  For instance, in the Coma Cluster (A1656), the
{\it ROSAT} PSPC image verifies the gravitational integrity of the southwestern
clump of galaxies, which is not clearly distinguishable in velocity space
(White
et al.\ 1993; Bird 1993).  In this sense X-ray and optical observations
complement each other well.  The ICM is a sensitive
tracer of the gravitational potential, but cannot reveal (with current
techniques) the presence of structure along our line-of-sight to a cluster.
Velocity data from optical or radio spectra can
distinguish systems superimposed along the line-of-sight, but yields little
information about subcluster mergers in the plane of the sky.
Therefore, in order to understand the dynamical state of any particular
cluster it is helpful to combine X-ray and optical observations.

A2151, the Hercules Cluster, is the prototypical rich, irregular cluster of
galaxies.  Its high spiral fraction, $\sim$50\%, makes it
an attractive target
for 21-cm.\ radio observations.  For example, Bird, Dickey \& Salpeter (1993,
hereafter BDS) supplement the optical catalog of redshifts using neutral
hydrogen observations of faint spirals.  BDS find evidence for a significant
level of substructure in A2151, verifying earlier detections by Geller \&
Beers (1982), Dressler \& Shectman (1988) and West \& Bothun (1990).  In
particular, BDS note that there is some contradiction between these optical
results, which find subclusters separated by about 17 arcminutes on the sky,
and the X-ray results of Magri et al.\ (1988; hereafter M88).  Their {\it
Einstein} IPC image of A2151 is clearly bimodal, but the peaks of X-ray
emission are separated by only about seven arcminutes.  The emission detected
by {\it Einstein} is confined to the central, highest density galaxy peak.
BDS proposed that the gas density in the other optically-identified
subclusters is too low to be detected by the IPC.

We have obtained a 12.5 ksec observation of A2151 with the {\it ROSAT}
PSPC.
The sensitivity of the PSPC to low
surface brightness features and its large field of view
enable us to directly
compare the large-scale morphology of the gas and galaxy distributions
in this spiral-rich irregular cluster.  The PSPC image reveals that, unlike
the case in Coma and most other
clusters, the gas and galaxy morphologies in A2151 are
very {\it dissimilar}.  The long integration
time also makes it possible to compare temperature and metallicity estimates
in
different regions of the cluster.

In Section 2, we present the reduction and analysis of the {\it ROSAT}
image and spectra.  In Section 3, we use the information derived from
the X-ray image to constrain the analysis of the optical and 21-cm redshift
catalog from BDS.  We have supplemented their velocity and position data
with morphological types taken from the Dressler (1980) catalog.
This additional information allows us to look for
variations in spiral fraction and kinematic differences between ellipticals
and spirals, which may correlate with properties of the ICM (M88;
Zabludoff \& Franx 1993).  In Section 4 we present a summary of the A2151
observations and our interpretation of its dynamical state.

\begin{center}
{\bf 2. X-RAY OBSERVATIONS \\}
\medskip
{\bf 2.1 Image Analysis}
\end{center}

The {\it ROSAT} PSPC (Position Sensitive Proportional Counter) is a wide-field
imaging instrument.  Its large field of view (2$^o$ across)
and
sensitivity to extended, low surface brightness emission make it an obvious
choice for imaging clusters of galaxies.  The PSPC has a FWHM of 30''
in the central 40 arcmin of the field of view, which is
equivalent to approximately 21$h^{-1}$ kpc at the recessional velocity
of the main Hercules subcluster (see below).  Here H$_0=75h$ km s$^{-1}$
Mpc$^{-1}$.  The resolution degrades toward the field edges to several
arcmin, which is about 100$h^{-1}$ kpc.  The front window of the
X-ray detector (which is protected by a fine wire mesh)
is supported by a ring and axis support. These supports block
incident X-rays and project a shadow onto the detector. This shadow can be
seen on the detector as a circular feature with a radius of 20\arcmin in the
center of the detector with 6 radial supports, which leaves the
central 40\arcmin of the detector clear of obstruction.
To smooth out the effects of the wire mesh, the satellite
is wobbled during the observation to provide even exposure over most of the
field of view.

We have performed our image and spectral reductions using the Post-Reduction
Offline Software (PROS) within the IRAF environment, and XSPEC to fit
spectral models.  In order to minimize the effects of solar X-rays, which
illuminate the detector after being scattered by the earth's atmosphere,
we first
screened the data using the total Accepted X-ray Events (AXE) count rate
as our discriminator. By smoothing the AXE and eliminating times where the
count rate exceeded 2$\sigma$ above the mean, we
can eliminate most of the scattered solar X-rays from our dataset.
This procedure eliminated about 400 seconds of the observing time,
or 3.2\% of the total time.
The A2151 image was corrected for vignetting using exposure
maps derived within the 7 energy bands of the PSPC as defined by
Snowden et al.\ (1993).
The effective
exposure maps in these 7 bands are generated using software at Goddard Space
Flight Center and
are based on the precise orientation of the satellite during the observation.
Having used these energy-dependent exposure maps for flatfielding, we then
co-add the images in the 0.5-2.0 keV energy range.  Excluding the low-energy
photons reduces contamination by galactic X-ray emission and reduces
the effects from
the long term enhancements (Snowden et al.\ 1992).
The background for the spatial analysis is taken
from our dataset using the northwestern
area of the image.  This region is an 11.5$^\prime$ circle centered at
$\alpha(J2000.0)=$16:02:55, $\delta(J2000.0)=$+18:02:40.3, well outside of the
central region of emission.
Visual inspection shows that this area is devoid of
cluster-related emission or point sources.  In the background region the
count rate is 0.3 counts/pixel, which in physical units is
3.96$\times$10$^{-4}$ counts/s/arcmin$^2$.
The flattened, background-subtracted image is smoothed with a 30''
Gaussian filter and shown in Figure 1.

Examination of Figure 1 reveals the presence of many point sources
coincident with the cluster.  In a forthcoming paper we will present an
analysis of these serendipitous sources, which for the most part are
coincident with optical galaxies.  We do not include them in the
rest of this study.

Comparison of the PSPC image with the corresponding optical image and
galaxy number density contours in Figure 2 (optical image provided by A.
Dressler; contour map taken from BDS)
reveals that in this
complicated galaxy cluster, the X-ray and optical distributions are {\it not}
very similar.  This is unlike either A1656 (White et al.\
1993) or A548 (Davis et al.\ 1993), two systems of different morphology
in which the
galaxies and ICM trace each other closely.  The bimodal structure seen in
the core of A2151 by M88 is clearly seen in the {\it ROSAT} image.
This complexity in the ICM is not obviously related to any similar structure
in the galaxy distribution, as comparison with Figure 2 reveals.

To verify that these sources are extended, we compared the
measured point spread function
of the PSPC to the profile for the sources.
To do this we extract the radial profile for the sources listed in
Table 1 in the energy range from 0.5 keV to 2.0 keV.  The profile was fit
using software included with the FTOOLS software package.  The modelled
PSF is a good fit for the measured profile for A2151N only.  To quantify
this, we list the measured half power radius for the sources in A2151.
The predicted half power radii are for point sources having the same spectral
distributions and off-axis angles as the A2151 sources.
We find
that both sources in the central region (designated as A2151C (bright) and
A2151C (faint) in Table 1) and A2151E have half power radii which are far
larger than the PSF:

\bigskip

\begin{tabular}{lcc}
Source & Half-power radius ('') & Predicted \\
A2151C (bright) & 101 & 24 \\
A2151C (faint) & 36 & 14 \\
A2151E & 195 & 30 \\
\end{tabular}

\bigskip
\noindent
On the other hand, the northern source is point-like, with an observed
half-power radius of 72'' and a predicted value of 66''.
Because the central and eastern sources are extended, we assume that
their emission is originating in the intracluster medium.

The
northern subcluster, first identified by BDS on the basis of its position
and kinematics, has no diffuse X-ray emission associated with it, at an
upper limit of 1.19$\times$10$^{-3}$ counts/s/arcmin$^2$ (the flux
limit is 1.2$\times$10$^{-14}$ erg/s/cm$^2$).
The point source in the northern subcluster has a power-law
spectrum (see next section) and is located at the position of NGC 6061.
NGC 6061 is listed as a radio-galaxy in the catalog of Zhao, Burns
\& Owen (1989).
We tentatively identify it as an active galaxy although it is not listed
in the AGN catalog of Veron-Cetty
\& Veron (1987).  We have been unable to find any
optical spectra of this galaxy to verify this identification.

The eastern galaxy enhancement, which was noted but
could not be verified by BDS on the
basis of its velocity distribution, does appear in the X-ray map.
Its spatial extent, the coincidence of its
position with the optical peak and its spectral characteristics (see next
section) suggest that it is physically associated with the subcluster.
This in turn suggests that the subcluster itself is physically distinct
from the central gas and galaxy peak.

\medskip

\begin{center}
{\bf 2.2 Spectral Analysis}
\end{center}

The PSPC spectral properties are unfortunately not quite as appropriate for
clusters as its imaging capabilities:  the PSPC detects individual X-ray
photons in the energy range 0.1-2.4 keV, with a resolution of ${\Delta E \over
E} \sim 0.4 ( {E \over {1 {\rm keV}}} )^{-1/2}$.
This energy range is somewhat lower than the
typical emission weighted mean temperature for rich clusters, 5-6 keV
(Mushotzky 1984).  The M88 observations of A2151 show that its ICM is
cooler and less luminous than is usual for a rich cluster, so the PSPC
nonetheless
should enable us to constrain the spectral properties of the Hercules system.

We select four areas of our A2151 image for spectral modelling:  the two peaks
seen near the cluster centroid (these were detected in the {\it Einstein}
image), the eastern emission, and the bright northern source.  We extracted
these regions from the broadband image using the PROS utility `qpspec'.
The contribution of unrejected charged particles was
modeled using the average master veto rate and removed from the extracted
spectra.
The background spectrum was extracted from the same region used in the image
analysis.
We used the XSPEC package (Shafer et al.\ 1990)
for the model fits.  For each of the 4 regions we tested
two different models, a Raymond-Smith plasma (for thermal emission from the
ICM) and a power-law spectrum (typical of active galactic nuclei).  In
addition, to verify that our spectral fits were not adversely affected by
the presence of more than one temperature component in the emission, for
the central regions we tested two-temperature models as well as the single
component description.  The individual fits are discussed below.

The relatively narrow energy bandpass and resolution of the
PSPC mean that, rather
than determining temperatures and metallicities directly from line ratios
as is done with optical spectroscopy, the {\it shape} of the X-ray
spectrum is used to estimate the physical properties of the system being
studied.  This strategy works well for cool clusters, in which the peak
temperature
is either within or at least close to the {\it ROSAT} bandpass.
For the brightest X-ray peak, a Raymond-Smith
plasma with temperature 1.67$^{+0.47}_{-0.25}$ keV and metallicity
0.56$^{+0.31}_{-0.20}$ solar produced a $\chi ^2$ = 24.92 for 19 degrees of
freedom.   (All confidence intervals quoted are 90\% for one
interesting parameter, with $\delta \chi ^2 = 2.71$.)
Here the galactic neutral hydrogen column density is held
constant at the Stark et al.\ (1992) value of $3.4 \times 10^{20}$ cm$^{-2}$.
This temperature determination is about 1 keV cooler than the M88 value
of 3 keV.  While M88 do not quote errors, typical
uncertainties in IPC temperature
determinations are 1.5 keV, so these values are marginally consistent.
The M88 {\it Einstein} observations did not permit accurate estimation of the
metallicity for comparison.  The X-ray luminosity of this component is
8.7$^{+9.7}_{-9.7}~\times$10$^{42} h^{-2}_{75}$ erg s$^{-1}$.
Fitting a two-temperature model does not improve the value of $\chi ^2$.

The secondary peak in the central X-ray emission has no obvious counterpart
in the corresponding galaxy distribution, although the S0/E galaxy NGC6047
is located at its center.
The best-fit model (with $\chi ^2$ = 14.21 for 19 d.o.f.) is
a thermal plasma with T=1.03$^{+0.60}_{-0.06}$ keV
and metallicity of 0.32$_{-0.10}^{+0.15}$ solar.
Adding either a second temperature component or a power-law component
does not reduce the value of $\chi ^2$ . Allowing the
value of the galactic absorbtion to be a free parameter does not change the
fitted parameters significantly.  The allowed range of the galactic column
density brackets the measured value.
Assuming that this feature is at the distance of A2151, the X-ray luminosity
in the 0.1 - 2.0 keV band is 3.4$^{+0.68}_{-0.61}~\times$10$^{42}
h^{-2}_{75}$ erg s$^{-1}$. This luminosity
is consistent with that of a poor cluster.
Therefore we interpret the secondary peak as a cooler, relatively distinct
component of the ICM.
Unfortunately the M88
observation of A2151 did not permit a derivation of temperature for this peak.

The diffuse eastern component of X-ray emission contributes about
300 photons in our PSPC image, enough to attempt to fit a model spectrum.
Once again we find that a single-temperature Raymond-Smith plasma provides
the most consistent fit to the observed spectrum, with
T=0.85$^{+0.21}_{-0.17}$ keV and A=0.16$^{+1.01}_{-0.13}$ solar.  The value
of $\chi ^2$ for this model is 14.6 with 19 d.o.f. This feature has a
luminosity of 5.2$^{+3.7}_{-3.2}~
\times$10$^{41} h^{-2}_{75}$ erg s$^{-1}$ in the 0.1 to 2.0
keV band of {\it ROSAT}, similar to that of a bright elliptical. However,
there is no elliptical located at the centroid of this
emission.  Note that this temperature is
significantly different from the temperature of the brightest peak.

Unlike the three brightest sources, the best-fit model for the northern X-ray
peak is a power law, with photon index 2.44$^{+0.18}_{-0.19}$.  Allowing the
neutral hydrogen column density to vary from the Stark et al.\ (1992) value
yields a range from 2.1-5.6$\times 10^{20}$ cm$^{-2}$, which is consistent
within the uncertainties.  The $\chi ^2$ for this model is quite good
(24.3 for 23 d.o.f.):  we tentatively identify the elliptical galaxy
at this position, NGC6061, as an active galaxy.  Assuming that the source
has this power-law spectrum and is located at the distance of A2151, it has
a luminosity of 1.6$^{+0.20}_{-0.20}~\times$10$^{41} h^{-2}_{75}$ erg s$^{-1}$.

The X-ray properties of these sources are summarized in Table 1.

\begin{center}
{\bf 3. OPTICAL ANALYSIS \\}
\medskip
{\bf 3.1 Quantifying the Substructure}
\end{center}

Unlike the apparently
more spherically-symmetric systems A1656 and A2256, no one has
ever doubted the presence of substructure in the Hercules Cluster.  Both
Dressler \& Shectman (1988) and West \& Bothun (1990) find significant levels
of substructure in A2151 using tests which combine velocity and position
information.  BDS use additional redshifts from optically-faint spirals,
as well as the deep position catalog generated with the Minnesota Automated
Plate Scanner (Dickey et al.\ 1987), to verify that the substructure is
present in the faint galaxies as well as the bright ones.  The kinematical
properties of the complete cluster dataset are provided in
Table 2.

We summarize the indicators of
substructure for the restricted, 126 member BDS
catalog in Table 3.  The restricted dataset includes galaxies which lie
within 3$S_{BI}$ of the cluster velocity (where $S_{BI}$ is the robust
estimator of dispersion described in Beers et al.\ 1990) and within 54$^\prime$
(or 2$R_C$ as defined by Tarenghi et al.\ 1979).  The diagnostics used are
defined as follows:  the skewness and kurtosis measure the symmetry and tail
population of the cluster velocity distribution relative to a Gaussian
(Fitchett \& Merritt 1988; Bird \& Beers 1993).  The Lee statistic (Fitchett
1988) quantifies the probability that the position data were drawn from a
two-group parent distribution rather than a single group.  The $\Delta$-,
$\alpha$-, and $\epsilon$-statistics (Dressler \& Shectman 1988; West \&
Bothun 1990; Bird 1993, 1994a respectively)
use velocity and position information
simultaneously to look for substructure.  These tests are described in more
detail in Bird (1993, 1994a)
 and references therein.  The substructure diagnostics
are normalized using a Monte Carlo resampling technique and 500 random
realizations.  By convention, in each of these
tests the null hypothesis is rejected if the significance level (the number
reported in Table 3) is less than 0.100.  That is, if less than 10\% of the
random realizations return a value of the test statistic which is greater
than or equal to the value given by the observed distribution,
we consider it a significant detection of
substructure.

The kurtosis and each of the three-dimensional statistics returns a positive
detection.  The value of the kurtosis (--0.563, where 0.000 is the
rescaled Gaussian
value) reveals that the velocity distribution is light-tailed compared to
Gaussian.  It has fewer members in the outer parts of the
distribution than expected for a Gaussian of the same dispersion.  This is
often the case when a distribution is composed of two or more distinct
subpopulations (Bird \& Beers 1993; Ashman, Bird \& Zepf 1994).
The velocity separation between the BDS northern
subcluster and the rest of the system is probably responsible for this
result.  (Note:  BDS find that their unrestricted dataset, which includes
3 high-velocity galaxies which do not survive the 3$S_{BI}$ filter, does
not possess a significantly non-Gaussian kurtosis, although a more robust
and conservative measure of tail population, the {\it tail index} TI (Bird
\& Beers 1993), is not consistent with Gaussian.  Obviously the handling of
outliers in the dataset can greatly influence the results of these tests,
particularly tests which are classical, moment-based estimators.)

The three-dimensional diagnostics are all sensitive to local correlations
between velocity and position.  Their significant rejections all suggest
the presence of kinematically-distinct subgroups in the A2151 dataset, which
is unsurprising given its complex velocity structure and irregular galaxy
distribution.  Malumuth et al.\ (1992) and Bird (1994b) discuss the effects
of smoothly-varying velocity fields and decreasing velocity-dispersion profiles
on the 3-D estimators.  Like A2107 (Oegerle \& Hill 1992), the average
velocity in A2151 varies relatively smoothly along its axis of elongation
(see BDS Figure 8).  Unlike A2107, however, the galaxy distribution in A2151
is extremely clumpy.  Therefore it seems unlikely that smooth rotation
explains the significant correlations between velocity and position in the
Hercules Cluster.

To allocate the individual galaxies in the BDS dataset to their host
subclusters, we have used the KMM mixture-modelling algorithm of McLachlan
\& Basford (1988).  KMM is an implementation of a maximum-likelihood
technique which assigns each galaxy into a prospective parent population,
and evaluates the improvement in fitting a multiple-group model over a
single-group model.  This evaluation is based on the {\it likelihood ratio
test statistic} or LRTS, which quantifies the probability that the given
dataset is consistent with a single-group null hypothesis.  KMM is applied
to velocity and position data for 25 galaxy clusters in Bird (1994a).

One major difficulty in applying any objective partitioning algorithm without
{\it a priori} knowledge of the system is the recognition and validation of
the number of
distinct groups present in the data.  Using a ``friends-of-friends''
technique based on the Dressler-Shectman $\Delta$-statistic, BDS find
statistical evidence for the existence of two groups, the central cluster
and the northern subcluster.  However, their technique cannot verify the
existence of the third peak seen in the optical contour map.  For
convenience, and for comparison with the {\it ROSAT} image, we provide the
BDS contour map superimposed on Dressler's image of A2151 in Figure 2.
This contour map was generated from the Automated Plate Scanner catalog of
A2151 published by Dickey et al.\ (1987).  Galaxies with magnitudes above
17.0 on the $E$-band Palomar Sky Survey plate (approximately equivalent
to a red bandpass)
are used to generate the map, to reduce contamination by background
galaxies.

Buoyed by the discovery of hot gas in the same position as the eastern
optical peak, and assuming that the kinematic evidence for the northern
subgroup is convincing, we require KMM to fit three groups to the
velocity and position data.  (The negative detection of structure by the Lee
statistic, in such an irregular system as A2151, also suggests that more than
two groups are present in the position data.)  Because the algorithm is so
sensitive to edge effects (i.e. the inclusion or exclusion of outliers can
greatly affect the parameters of the mixture model as noted above), we
have restricted the
analysis to galaxies meeting the redshift cutoff defined above, and also
lying within 33$^\prime$ or about 1 Abell radius
($\sim 1.5 h^{-1}$ Mpc) of the cluster centroid.
This eliminates 15 galaxies from the fit; they are in the extreme
north and extreme south of the system.  To verify that the substructure
detected above is not due to the presence of these outlying galaxies, we
summarize the substructure diagnostics in the radially-restricted dataset
in the second line of Table 3.  The kurtosis and the
nearest-neighbor
diagnostics still reject their null hypotheses at a significant level,
thereby showing that the structure in A2151 is present even in the core of
the system.
The LRTS for this subset is significant at a
level of $<$0.001, providing us with statistical evidence that the three
groups seen in the optical contour map are physically distinct.
That is, the KMM algorithm rejects the single group hypothesis for the A2151
dataset at a confidence level of $>$ 99.9\%.

We have used the posterior probabilities assigned by the KMM algorithm to
allocate each galaxy in the (radially-restricted) dataset to its host
subcluster.  The kinematic properties of the subsystems, coded A2151C,
A2151N, and A2151E (for central, northern and eastern, respectively) are
compared to each other and to the global cluster properties in Table 3.
The position centroids are indicated by the letters C, N and E in Figure
2.  Note that the use of velocity information in the KMM partition has
skewed the centroid of A2151E and A2151N
from the position dictated by the galaxy
distribution alone.
The coincidence between the positions of the X-ray and optical peaks, for
A2151C and A2151E, provides some verification that KMM has allocated galaxies
sensibly.  (These positions are summarized in Table 4 and discussed in
further detail in the next section.)  In Figure 3 we present the velocity
histograms for the three subclusters.

We check the partition more closely by looking for additional
indications of substructure within the subclusters themselves.  This is
somewhat risky because they contain so few galaxies, but
the resampling technique used for
determining significance levels for all the substructure tests accurately
estimates the uncertainty in the test statistic due to Poisson fluctuations,
even in small datasets.  Unlike the complete dataset or the
radially-restricted subset, for which several of the substructure diagnostics
were positive, only A2151C shows any indication of substructure.  The
central subcluster has a significant $\Delta$-statistic although none of the
other tests are positive.  A2151N and A2151E possess no detectable
substructure.  This result suggests that KMM has divided the cluster into
reasonable,
if not unique subsystems.

\begin{center}
{\bf 3.2 Kinematical and Morphological Properties of the A2151 Subclusters}
\end{center}

We have supplemented the kinematical and dynamical information with the
individual galaxy morphological types taken from Dressler (1980).
Many authors have noted
that a cluster's spiral fraction is
anticorrelated with the presence of hot intracluster gas
(Bahcall 1977; M88; Arnaud et al.\ 1992).  The spiral
fractions (not including S0's) for the 3 subclusters are A2151C:  49\%
(30 out of 61); A2151N:  63\% (20 out of 32); A2151E:  53\% (8 out of 15).
In each of the subclusters one galaxy was not included in the Dressler
catalog and has therefore been removed from further morphological analysis.
The spiral fractions are all consistent within the sampling errors despite
the differences in gas properties in the three subsystems.  High-resolution
VLA 21-cm maps of spirals in A2151 (Dickey 1995) reveal
that the spirals in A2151C and A2151E are considerably deficient in
neutral hydrogen compared to spirals in the north.  This is presumably
related to the lack of intracluster gas in the north,
either as an agent of ram-pressure
stripping or as a repository for the missing gas.

Even when the substructure is considered,
however, kinematical differences between the spirals, ellipticals and S0's
exist.  These properties of the morphological subsets are presented
in Table 5.  In A2151C the spirals have a higher average velocity
(defined using the robust estimator $C_{BI}$; Beers, Flynn \& Gebhardt 1990)
than the ellipticals. In A2151N the offset exists but is only marginally
significant.  (We have not considered A2151E in this discussion because it
contains only 2 ellipticals.)  In A2151N the two subsets have similar
velocity scales (the velocity dispersion or scale is defined using the
robust estimator $S_{BI}$).
In A2151C, however, the three subsets are quite different.  The velocity
locations of the spirals is much different from that of the earlier types.
This may be an indication of spiral infall, especially since their velocity
location is higher than that of the ellipticals and S0's (that is, it is
more similar to the velocity of A2151N which may be a source of galaxies).
More difficult to understand, however, is the fact that the velocity
dispersion for the S0 galaxies
is much higher than that of either the ellipticals or spirals.  This effect is
only
evident when the cluster's structure is considered, so it was not obvious in
the Zabludoff \& Franx (1993) results.  They do find a similar situation in
DC2048-52, a cluster with smoother galaxy distribution than A2151 but with
strong evidence for substructure from the nearest-neighbor diagnostics
(Bird 1993).

\medskip

\begin{center}
{\bf 4.  SUMMARY AND INTERPRETATION}
\end{center}

We combine our conclusions about the X-ray and optical substructure
in the Hercules Cluster to suggest a consistent model of this cluster's
dynamical state.

1)  Our PSPC image reveals that the BDS northern subcluster has no detectable
diffuse X-ray emission.  The eastern peak in the galaxy distribution which
BDS could not verify does coincide with an  enhancement above the X-ray
background, suggesting that it is physically distinct.  The bimodality that
M88 detected in the central cluster emission does not correspond to any
feature in the galaxy distribution in the same area.

2)  Analysis of the X-ray spectra reveal a difference in temperature
between A2151C and A2151E.
A2151E is significantly cooler
than the brightest peak of emission (0.85 keV vs.\ 1.67 keV),  although it is
marginally consistent with the second central peak.  The complexity in the
X-ray structure, both spatial and temperature, implies that the ICM in the
Hercules Cluster is far from its equilibrium state.

3)  A maximum-likelihood analysis of the BDS galaxy velocities and positions,
constrained to fit three subclusters, reveals significant kinematic
differences between A2151C, A2151N and A2151E.  A2151N is
higher in line-of-sight velocity than either of its
companion systems.  The subset of S0 galaxies in A2151C has a much higher
velocity scale than either the spirals or ellipticals.  The velocity
dispersion of A2151E is consistent (with the 90\% confidence intervals)
with that of A2151C, although its ICM is significantly cooler.

The low temperatures of the diffuse cluster emission move A2151 even farther
away from the $\sigma _v - T_x$ relationship found by Edge \& Stewart
(1991) (see also Bird \& Mushotzky 1994)
than {\it Einstein} results suggested.  This relation, derived from
virial considerations, relates the properties of the galaxies to those of
the gas:
\begin{equation}
T_x = 6.026 \times 10^{-4} \sigma _v^{1.35}
\nonumber
\end{equation}
where T$_x$ is in units of keV and $\sigma_v$ is in units of \ksec .
Equation (1)  predicts that A2151C should have an X-ray temperature of 4.2 keV,
far
warmer than the PSPC value.  A2151E is predicted to be at
3.2 keV, much hotter than the 0.8 keV gas we detect in the PSPC image.
Similarly, A2151N, which has no detectable emission, is predicted to have
a temperature of 2.3 keV.  We take this disagreement as evidence
of a recent merger event in A2151, which has disrupted the (assumed)
hydrostatic equilibrium of the ICM within the subclusters.
The presence of substructure in both the galaxy and gas distributions is
clear in A2151.  If the cluster is pre-merger, the subclusters should
themselves be virialized and unperturbed in most hierarchical formation models
(as is the case in the well-known
bimodal system A548; Davis et al.\ 1994).  Even within the subclusters,
however,
the virial condition defined above is violated.  We suggest that this is
because the subclusters have been violently disrupted by their interaction.

Because even
within the subclusters the galaxies and gas are out of equilibrium with
each other, mass estimation based on either galaxy or gas dynamics (for which
an equilibrium situation must be assumed) is highly inaccurate.
For this reason we have not derived either an X-ray or
an optical mass for any of the A2151 subclusters.

Burns et al.\ (1994, hereafter B94) find evidence using {\it ROSAT} images
of clusters that radio galaxies, especially wide-angle tail sources (WATs),
are preferentially located within clumps of X-ray gas.  They interpret the
existence of the WAT as evidence of extreme turbulence in the ICM, i.e. as
yet another indication of a recent merger event.  Specific examples of this
phenomenon can be seen in A400 (Beers et al.\ 1990) and
A2634 (Pinkney et al.\ 1993), which both have extremely radio-loud
central galaxies.
To test this hypothesis
for A2151, we have used the catalog of Zhao et al.\ (1989)
to look for WAT galaxies in A2151.  In Table 4, the
results of this search are compared to the X-ray and optical peaks from the
present study.  A2151E does not contain an identified radio source, but
both of the central peaks as well as the northern active galaxy do coincide
with radio galaxies.  NGC 6040, located at the brightest X-ray peak, is a
narrow
angle tail galaxy.  Both NGC 6047, which coincides with the secondary central
peak, and the previously-mentioned NGC 6061 (see Sect.\ 2) are WAT sources.
The lack of detectable gas in A2151N provides some difficulty for a recent
merger as a trigger of WAT activity in NGC 6061, but is consistent with the
25\% of sources found by B94 which are not correlated with any obvious X-ray
emission.

Computer simulations of subcluster mergers also presented by B94 lend
further weight to the post-merger interpretation of A2151C and A2151E.
Their code, described in detail in Roettinger, Burns \& Loken (1993),
includes N-body and hydrodynamical algorithms to follow evolution of dark
matter, galaxies and gas, as well as both bremsstrahlung and line-emission
cooling.  They find that the less-massive subcluster is ram-pressure stripped
of {\it most} of its gas, and that the remaining gas is cooler, similar
to the observed situation in A2151C and A2151E.  In this interpretation,
the secondary X-ray emission centered on NGC 6047 is the ram-pressure
stripped gas from A2151E, trapped within the presumably deeper gravitational
potential of A2151C.

Determining whether any particular cluster is pre- or post-merger is an
extremely difficult
(if not impossible) task, and we would therefore like to stress that our
assessment of A2151 as a post-merger system is tentative.
Nonetheless, we believe that the following observations lend support to
this interpretation.  The deviation of the A2151 subclusters (A2151C and
A2151E) from the virial relationship between galaxy velocity dispersion and
X-ray temperature suggests that some disruptive physical event must have
already occurred.  If the subclusters had not interacted, one might expect
that their gas and galaxies would be in agreement with the
$\sigma _v - T_x$ relationship, as is the case in A548 (Davis et al.\ 1994).
(Alternatively, it is possible that A2151C and A2151E are themselves in
the process of formation.)
Similarly, B94 argues that the presence of WAT radio
sources in clusters is a signal that subcluster mergers have recently
occurred.  The radio activity of NGC 6040 and NGC 6047 may therefore
indicate that A2151C has just undergone a merger event.

We interpret the
lack of X-ray gas corresponding to the optical galaxy enhancement A2151N
to mean that those galaxies are not themselves a physically-distinct
subgroup, but are instead galaxies infalling from the Hercules Supercluster.
In general, spiral infall is detected in clusters when late-type galaxies
are preferentially located in the outer parts of the cluster, and when
they possess a much larger velocity dispersion than early-type galaxies
(cf.\ Virgo, Tully \& Shaya 1984; A2634, Scodeggio et al.\ 1994).  Such
an enhanced velocity dispersion is not seen in any of the A2151 subclusters.
Nonetheless, because A2151N lies within the Hercules Supercluster, it is
possible that its velocity dispersion is lower than might be expected for a
system embedded within a less structured environment.

The combination of imaging and spectroscopy in X-ray, optical and radio
wavelengths, as well as physically realistic computer simulations, provides
us with a possible understanding of A2151, the Hercules Cluster.  In this
complex
system, we find evidence for a recent merger event as well as galaxy infall
into the cluster core.  Each of the observational and computational
components provides an important part of the dynamical analysis of this
cluster of galaxies.  The value of a multiwavelength approach to galaxy
clusters can hardly be over-emphasized.

\medskip
\medskip

The preliminary X-ray analysis of A2151 was performed at the {\it ROSAT}
Guest Observers' Facility at Goddard Space Flight Center.  CMB would
especially like to thank Karen Smale, Eric Schlegel and Mike Corcoran for
their hospitality and efficiency.  It is a pleasure to thank Alan Dressler
for use of his detailed photographic plate of the cluster.  We would like to
thank Ann Zabludoff and Ray White for conversation and
reassurance during the course of this project.  CMB is also grateful to Jack
Burns, Jason Pinkney, Kurt Roettinger and the Astronomy Department at New
Mexico State University for two days of extremely helpful discussions.
Richard Mushotzky greatly improved the text through his critical reading.
This research was supported in part by NASA Grant No.\ NAG5-2434
to Michigan State University and
NSF Grant No.\ OSR-9255223 to the University of Kansas.

\newpage
\begin{center}
{\bf REFERENCES}
\end{center}

\ref{Arnaud, M., Rothenlug, R., Boulade, O., Vigroux, L \& Vangioni-Flam, E.
1992, A\&A, 254, 49}
\ref{Ashman, K.M., Bird, C.M. \& Zepf, S.E.  1994, AJ, submitted}
\ref{Bahcall, N.A.  1977, ApJL, 218, L93}
\ref{Baier, F.W.  1983, Astron. Nacht., 5, 211}
\ref{Beers, T.C., Forman, W., Huchra, J.P., Jones, C. \& Gebhardt, K.
1991, AJ, 102, 1581}
\ref{Beers, T.C., Flynn, K. \& Gebhardt, K.  1990, AJ, 100, 32}
\ref{Bird, C.M.  1993, Ph.D thesis, University of Minnesota and Michigan
State University}
\ref{Bird, C.M.  1994a, AJ, 107, 1637}
\ref{Bird, C.M.  1994b, ApJ, 422, 480}
\ref{Bird, C.M. \& Beers, T.C.  1993, AJ, 105, 1596}
\ref{Bird, C.M., Dickey, J.M. \& Salpeter, E.E.  1993, ApJ, 404, 81 (BDS)}
\ref{Bird, C.M. \& Mushotzky, R.F.  1994, ApJ, submitted}
\ref{Burns, J.O., Rhee, G., Owen, F.N. \& Pinkney, J.  1994, ApJ, to appear
1 March}
\ref{Burns, J.O., Roettiger, K., Pinkney, J., Loken, C., Doe, S., Owen, F.,
Voges, W. \& White, R.  1994, to appear in the {\it Proceedings of the ROSAT
Science Symposium} (B94)}
\ref{David, L.P., Slyz, A., Jones, C., Forman, W., Vrtilek, S.D. \&
Arnaud, K.A.  1993, ApJ, 412, 479}
\ref{Davis, D. S. \& Mushotzky, R. F. 1993, AJ, 105, 491}
\ref{Davis, D. S., Bird, C. M., Mushotzky, R. F. \& Odewahn, S. C. 1993, ApJ
sub
mitted}
\ref{Dickey, J.M., Keller, D.T., Pennington, R. \& Salpeter, E.E.  1987,
AJ, 93, 788}
\ref{Dressler, A.  1980. ApJS, 42, 565}
\ref{Dressler, A. \& Shectman, S.  1988, AJ, 95, 985}
\ref{Edge, A.C. \& Stewart, G.C.  1991, MNRAS, 252, 428}
\ref{Fitchett, M.J.  1988, MNRAS, 230, 169}
\ref{Fitchett, M. \& Merritt, D.  1988, ApJ, 335, 18}
\ref{Forman, W., Bechtold, J., Blair, R., Giacconi, R., Van Speybroeck, L.,
\& Jones, C.  1981, ApJLett, 243, L133}
\ref{Geller, M.J. \& Beers, T.C.  1982, PASP, 94, 421}
\ref{Henry, J.P., Henriksen, M., Charles, P., \& Thorstensin, J. 1981, ApJL,
243, L137}
\ref{Jones, C. \& Forman, W.  1992, in {\it Clusters and Superclusters
of Galaxies},
ed. A.C. Fabian, (Dordrecht:  Kluwer), 49}
\ref{Kent, S.M. \& Gunn, J.E.  1982, AJ, 87, 945}
\ref{Kent, S.M. \& Sargent, W.L.W.  1983, AJ, 88, 697}
\ref{Magri, C., Haynes, M., Forman, W., Jones, C., \& Giovanelli, R.
1988, ApJ, 333, 136 (M88)}
\ref{Malumuth, E.M., Kriss, G.A., Van Dyke Dixon, W., Ferguson, H.C.
\& Ritchie, C.  1992, AJ, 104, 495}
\ref{McLachlan, G.J. \& Basford, K.E.  1988, {\it Mixture Models}, (New York:
Ma
rcel Dekker)}
\ref{Mohr, J., Fabricant, D. \& Geller, M.  1993, CfA preprint}
\ref{Mushotzky, R.F.  1984, Physica Scripta, T7, 157}
\ref{Roettinger, K., Burns, J. \& Loken, C.  1993, ApJ, 407, L53}
\ref{Scodeggio, M., Solanes, J.M., Giovanelli, R. \& Haynes, M.P.  1994,
Cornell preprint}
\ref{Snowden, S. L., McCammon, D., Burrows, D. N. \& Mendenhall, J. A.
1993 ApJ, submitted}
\ref{Stark, A.A., Gammie, C.F., Wilson, R.W., Bally, J., Linke, R.A.,
Heiles, C.M. \& Hurwitz, M.  1992, ApJSuppl, 79, 77}
\ref{Tarenghi, M., Tifft, W.G., Chincarini, G., Rood, H.J. \& Thompson, L.A.
1979, ApJ, 234, 793}
\ref{Tully, R.B. \& Shaya, E.  1984, ApJ, 281, 31}
\ref{Veron-Cetty, M.-P. \& Veron, P. 1987, ESO Scientific Report \#5}
\ref{West, M.J. \& Bothun, G.D.  1990, ApJ, 350, 36}
\ref{White, S.D.M.  1992, in {\it Clusters and Superclusters of Galaxies},
ed. A.C. Fabian, (Dordrecht:  Kluwer), 17}
\ref{White, S.D.M., Briel, U.G. \& Henry, J.P.  1993, MNRAS, 261, L8}
\ref{Zabludoff, A.I. \& Franx, M.  1993, AJ, 106, 1314}
\ref{Zhao, J-H., Burns, J.O. \& Owen, F.N.  1989, AJ, 98, 64}

\newpage

\baselineskip 12pt
\begin{center}
T{\small ABLE} 1.  X-ray Properties of Bright Sources in PSPC Image.
\medskip
\medskip
\begin{tabular}{lcccccc} \hline \hline
Source & Extraction & Flux & L$_{0.1 - 2.0}$ & kT & A & $\chi ^2 / \nu$ \\
 & Radius & (10$^{-12}$ ergs & (10$^{42}~h^{-2}$ & (keV) & (solar units) & \\
 & ($h^{-1}$ kpc) & s$^{-1}$ cm$^{-2}$) &
ergs s$^{-1}$) & & & \\
\hline
A2151C (bright) & 210 & 3.1 $^{+0.26}_{-0.26}$
& 8.7$^{+9.7}_{-9.7}$ & 1.67$^{+0.47}_{-0.25}$ &
0.56$^{+0.31}_{-0.20}$ & 24.92$/$19 \\
A2151C (faint) & 200 & 1.3$^{+0.24}_{-0.22}$ & 3.6$^{+0.68}_{-0.61}$
 & 1.03$^{+0.60}_{-0.06}$ & 0.32$^{+0.15}_{-0.10}$
& 14.21$/$19 \\
A2151E & 130 & 1.9$^{+1.33}_{-1.81}$ & 0.050$^{+0.37}_{-0.32}$
 & 0.85$^{+0.21}_{-0.17}$ & 0.16$^{+1.01}_{-0.13}$
& 14.6$/$19 \\
A2151N & 151 & $<$ 1.2 $\times$ 10$^{-14}$ & & & & \\
\hline
 & & & & $\alpha$ & & \\
\hline
NGC 6061 & 151 & 0.57$^{+0.07}_{-0.07}$
& 1.6$^{+0.20}_{-0.20}$ & 2.44$^{+0.18}_{-0.19}$ & & 24.3$/$23 \\
\hline \hline
\end{tabular}
\end{center}

Notes to T{\small ABLE} 1.
The positions of the {\it ROSAT} sources are provided in T{\small ABLE} 4
for easier comparison with their optical and radio counterparts.

For all spectral models, the galactic hydrogen abundance was held constant
at the Stark et al.\ (1992) value of $3.4 \times 10^{20}$ cm$^{-2}$.
See text for further discussion.

\newpage

\renewcommand{\arraystretch}{1.1}

\begin{center}
T{\small ABLE} 2.  Dynamical Properties of A2151 and Subclusters.
\medskip
\medskip
\begin{tabular}{llcc} \hline \hline
System & $\alpha (J2000.0)$ & $C_{BI}$ (km s$^{-1}$) & $S_{BI}$ (km s$^{-1}$)
\\

& $\delta (J2000.0)$ & $IC_{BI}$ (90\%) & $IS_{BI}$ (90\%) \\
\hline
A2151 & 16:05:37.2 & 11066 & 752 \\
 & 17:49:08.4 & (--121, 116) & (684, 826) \\
A2151$^{\prime}$ & 16:05:25.9 & 11070 & 766 \\
 & 17:47:50.3 & (--131, 115) & (696, 852) \\
A2151C & 16:05:15.5 & 10650 & 707 \\
 & 17:39:45 & (--153, 152) & (623, 820) \\
A2151N & 16:05:55.0 & 11445 & 455 \\
 & 18:08:27 & (--123, 138) & (384, 549) \\
A2151E & 16:06:44.2 & 11756 & 570 \\
 & 17:46:13 & (--247, 437) & (412, 820) \\
\hline \hline
\end{tabular}
\end{center}

Note to T{\small ABLE} 2.
\noindent
A2151$^{\prime}$ designates the radially-restricted dataset used in the
KMM partition.

\newpage
\renewcommand{\arraystretch}{1.1}

\begin{center}
T{\small ABLE} 3.  Substructure Diagnostics for A2151 and Subclusters.
\medskip
\medskip
\begin{tabular}{lcccccc} \hline \hline
System & Skewness & Kurtosis & Lee & $\Delta$ & $\alpha$ & $\epsilon$ \\
\hline
A2151 & 0.242 & 0.061 & 0.134 & $<$0.001 & 0.012 & $<$0.001 \\
A2151$^\prime$ & 0.216 & 0.072 & 0.558 & $<$0.001 & 0.034 & $<$0.001 \\
A2151C & 0.160 & 0.344 & 0.120 & 0.002 & 0.128 & 0.168 \\
A2151N & 0.355 & 0.338 & 0.736 & 0.364 & 0.182 & 0.388 \\
A2151S & 0.160 & 0.344 & 0.946 & 0.142 & 0.502 & 0.584 \\
\hline \hline
\end{tabular}
\end{center}

Note to T{\small ABLE} 3.
\noindent
We have quoted significance levels for each of the substructure diagnostics.
By convention, substructure is detected if this number is less than 0.100.

\medskip
\medskip

\newpage

\begin{center}
T{\small ABLE} 4.  X-ray, Radio and Optical Correspondences
\medskip
\medskip
\begin{tabular}{lccc} \hline \hline
 & {\it ROSAT} peak & Radio Source & Optical ID \\
\hline
A2151C & & & \\
\hline
$\alpha (J2000.0)$ &16:04:33.9 & 16:04:26.4 & NGC 6040$^\dagger$ \\
$\delta (J2000.0)$ &17:43:13.6 & 17:44:30.6 & \\
\hline
$\alpha (J2000.0)$ &16:05:07.5 & 16:05:09.0 & NGC 6047$^\dagger$ \\
$\delta (J2000.0)$ &17:43:43.9 & 17:43:47.4 & \\
\hline
A2151E & & & \\
\hline
$\alpha (J2000.0)$ &16:06:33.6 & no radio & 16:06:44.3$^{\dagger \dagger}$ \\
$\delta (J2000.0)$ &17:43:58.3 & counterpart & 17:46:12.6 \\
\hline
A2151N & & & \\
\hline
$\alpha (J2000.0)$ &no diffuse & no central & 16:05:55.0 \\
$\delta (J2000.0)$ &X-ray component&radio source &18:08:27.1 \\
\hline
$\alpha (J2000.0)$ &16:06:14.9 & 16:06:16.3 & NGC 6061 \\
$\delta (J2000.0)$ &18:15:13.6 & 18:14:59.5 & \\
\hline \hline
\end{tabular}
\end{center}

\noindent
$^\dagger$ Both NGC 6040 and NGC 6047 are within the 1$\sigma$ position
centroids of A2151C.

\noindent
$^{\dagger \dagger}$ The positions listed for A2151E and A2151N are
taken from the KMM partitions.

\newpage

\begin{center}
T{\small ABLE} 5.  Kinematic Properties of the Morphological Sample
\medskip
\medskip
\begin{tabular}{lcc} \hline \hline
Subset (N$_{gal}$) & $C_{BI}$ (km s$^{-1}$) & $S_{BI}$ (km s$^{-1}$) \\
\hline
 & A2151C & \\
\hline
Ellipticals (10) & 10172 (--226, 230) & 404 (294, 556) \\
S0's (19) & 10547 (--404, 395) & 882 (732, 1127) \\
Spirals (27) & 10793 (--189, 195) & 573 (448, 764) \\
\hline
 & A2151N & \\
\hline
E+S0's (12) & 11305 (--234, 217) & 453 (323, 632) \\
Spirals (19) & 11568 (--244, 119) & 455 (324, 565) \\
\hline \hline
\end{tabular}
\end{center}

\newpage

\begin{center}
{\bf FIGURE CAPTIONS}
\end{center}

\ref{Figure 1.  The {\it ROSAT} PSPC image of A2151.}

\ref{Figure 2.  Dressler's Schmidt image of A2151 (Dressler 1980).
The overlaid contour map is taken from BDS.  The contours are linearly
spaced in units of 0.0257 galaxies arcmin$^{-2}$ and range from 0.0015
to 0.2582 galaxies arcmin$^{-2}$ from lowest to highest.  The letters C,
N and E indicate the position centroids of the groups identified by KMM
(see Section 3.1 for discussion).}

\ref{Figure 3.  Velocity histograms for A2151C (top panel), A2151N (middle
panel), and A2151E (bottom panel).}

\end{document}